\documentclass[prd,aps,twocolumn,superscriptaddress,preprintnumbers,showpacs,nofootinbib,tightenlines,floatfix,amssymb]{revtex4}
\usepackage{epsfig}

\newcommand{\beqn}{\begin{eqnarray}}
\newcommand{\eeqn}{\end{eqnarray}}
\newcommand{\eq}[1]{(\ref{#1})}

\newcommand{\w}{{\mathrm {wr}}}
\newcommand{\phys}{{\mathrm {phys}}}

\newcommand{\Z}{{Z \!\!\! Z}}
\newcommand{\perc}{{\mathrm{perc}}}
\def\bbbone{{\mathchoice {\rm 1\mskip-4mu l} {\rm 1\mskip-4mu l}
{\rm 1\mskip-4.5mu l} {\rm 1\mskip-5mu l}}}
\begin{document}

\title{Deconfinement phase transition in mirror of symmetries}

\author{M.N.~Chernodub}
\affiliation{Laboratoire de Mathematiques et Physique Theorique,
CNRS UMR 6083, F\'ed\'eration Denis Poisson, Universit\'e de Tours,
Parc de Grandmont, F37200, Tours, France}
\affiliation{Department  of Mathematical Physics and Astronomy,
University of Gent, Krijgslaan 281, S9, B-9000 Gent, Belgium}
\affiliation{Institute for Theoretical and
Experimental Physics, B.~Cheremushkinskaya 25, Moscow, 117218, Russia}
\author{Atsushi Nakamura}
\address{Research Institute for Information Science and Education,
Hiroshima University, Higashi-Hiroshima, 739-8527, Japan}

\author{V.I.Zakharov}
\affiliation{Institute for Theoretical and
Experimental Physics, B.~Cheremushkinskaya 25, Moscow, 117218, Russia}
\affiliation{Max-Planck-Institut f\"ur Physik, F\"ohringer Ring 6, 80805 M\"unich, Germany}

\preprint{ITEP-LAT/2009-01}

\begin{abstract}
We summarize and extend evidence that the deconfinement phase transition
in Yang-Mills theories can be viewed as change of effective non-perturbative
degrees of freedom and of symmetries of their interactions. In short, the strings
in four dimensions (4d)
at temperatures below the critical temperature $T_c$ are replaced by
particles, or field theories in 3d at $T>T_c$.  The picture emerges
within various approaches
based, in particular, on dual models,  lattice data and
field theoretic models. We concentrate mostly on the lattice data,
or on the language of quantum geometry.
\end{abstract}


\date{April 06, 2009}

\maketitle

\section{Introductory remarks}

Non-perturbative phenomena in Yang-Mills theories attract attention of theorists
since long. The most famous problem is  the confinement of color.
More recently, observation of the 'strongly interacting quark-gluon plasma'
(for review see, e.g. \cite{review})
fueled the fascination.

Quasiclassical solutions seem to be most natural starting point to
consider non-perturbative phenomena.
And, indeed, instantons provide a clue to understand generically
spontaneous chiral symmetry breaking.
%(although constructing detailed models is
%not straightforward because of the infrared uncertainties).
In contrast, there are no classical solutions of Yang-Mills equations which could
be relevant to confinement.

String picture is best suited to approach the confinement.
The deconfinement phase transition is then transition to percolation
of strings \cite{polyakov}. In terms of the Yang-Mills fields the first stringy
operator with well established relation to deconfinement is the Polyakov line,
\begin{equation}
L~=~P~\exp\Big\{-i\int _0^{1/T}A_0\,d \tau \Big\}\,.
\end{equation}
The vacuum expectation value of the Polyakov line is an order parameter
for the confinement-deconfinement phase transition.
Moreover, there are widely discussed models, see, e.g., \cite{yaffe,pisarski,kurkela}
which assume that the Polyakov line becomes a dynamical variable
with potential (in its simplest form):
\begin{equation}\label{potential}
V(\langle Tr L\rangle)~=~c_2\langle Tr L\rangle^2+c_4\langle Tr L\rangle^4~,
\end{equation}
so that the deconfining phase  corresponds to the tachyonic sign of the coefficient $c_2$.

Phenomenologically, confinement and chiral symmetry breaking are  strongly correlated and 
it seems unsatisfactory to explain these phenomena within
models which are not related to each other. A framework to reach synthesis
of instantons and, say, Polyakov's lines
is provided by dual models, or gauge/string duality, for review see,
 e.g. \cite{aharony}. Within these models all the non-perturbative effects
 of gauge theories are related to properties of strings and/or D-branes
 in extra dimensions.

 The model \cite{sakai} turned to be most successful to unify phenomenology of confinement and of chiral symmetry breaking. The geometry of the model introduces, in particular,
 a confinement-related horizon in an extra coordinate $u$ and a
 compact coordinate $x_4$:
 \begin{eqnarray}\label{extradimension}
  ds^2~=~\Big({u\over R_0}\Big)^{3/2}\big(-dt^2+\delta_{ij}dx^idx^j+f(u)dx^2_4\big)~+~\\\nonumber
  \Big({u\over R_0}\Big)^{-3/2}\Big({du^2\over f(u)}+u^2d\Omega^2_4\Big)~~,
  \end{eqnarray}
  where
  $$f(u)~=~1-\Big({u\over u_{\Lambda}}\Big)^3~~$$
  and $u_{\Lambda}$ is the position of the horizon, $d\Omega_4^2$ is the
 metric of a four-dimensional sphere; the corresponding coordinates will not
 play any crucial role in our considerations.
 At non-zero temperature, $T\neq 0$ there are two compact coordinates, $x_4$ and
 the Euclidean time $\tau$. The deconfinement phase transition is the interchange
 of geometry in the coordinates $x_4$ and $\tau$.

 As is emphasized in \cite{gorsky} the geometrical picture outlined above
 reproduces main features of the phenomenology, such as condensation
 of magnetic degrees of freedom at $T=0$ or Polyakov's lines becoming tensionless
 at $T=T_c$. Moreover, it predicts existence of the magnetic component
 of plasma which might be crucial to explain the unusual properties of the
 plasma
 \cite{chernodub,shuryak,nakamura,sasha,massimo}. Transition at $T=T_c$
 from 4d strings to 3d field theories, mentioned in the abstract, is predicted
 as well.

 Although the gauge/string duality suggests an exciting perspective of a unified
 description of the confinement and of strongly interacting plasma,
 this description is mostly qualitative at the moment. Indeed,
 conclusions mentioned above are based exclusively on consideration of topology
 of the extra dimensions (\ref{extradimension}). Exact metric (\ref{extradimension})
 is a property of large $N_c$-approximation and can hardly be true in the realistic
 case. Also, theory predicts a strong first order phase transition while it
 is either second order in case of SU(2) gauge group or weak first order in case
 of SU(3).

Thus, the gauge/string duality considerations would remain pure speculations
unless they can be complemented by other, more quantitative evidence.
In this note we turn to the lattice evidence concerning
the magnetic defects at $T\neq 0$
(for a review of the $T=0$ case see \cite{zakharov}). The language used by the lattice is in fact the language of quantum geometry, see, e.g., \cite{polyakov3}.
The main result of the present paper is that the lattice  data indeed demonstrate
transition from strings living in 4d at $T<T_c$ to 3d field theories at $T>T_C$.
Moreover, the lattice data allow to establish some particular features of the
emerging 3d theories, such as Higgs condensation of a scalar field.

Another quantitative approach to the deconfinement is the use of effective
theories. We already mentioned a prototype of such theories, see Eq (\ref{potential}).
Nowadays there exist much more developed versions of course. In particular, quite
a detailed comparison of the lattice-motivated models  with the analytical model
of Ref. \cite{kurkela} turns possible. In their gross features the models turn similar
but there exist important differences as well.

\section{Three-dimensional magnetic Higgs field on the lattice}
\subsection{Percolation of 1d defects}

Lattice data which we will analyze refers to 'defects' which are nothing else
but (closed) lines  and (closed) surfaces.
Both the lines and surfaces are defined as infinitely thin.
Historically, the defects were identified empirically
as configurations responsible for confinement. The 1d defects are
called magnetic monopoles while surfaces are center vortices, for review and
further references see \cite{greensite}. Originally, the defects were
defined in specific lattice language of the so called projected fields.
Later, they were identified in the language used by the continuum theory
\cite{zakharov}.
In particular, the surfaces are identified as magnetic vortices, or strings.
By definition, the magnetic strings
are closed surfaces in the (Euclidean) vacuum which can be open
on an external 't Hooft line.

We will not go into all the details now and only emphasize
that the language appropriate for the lattice data is that of the quantum
geometry. As a simplest example of this type let us remind the reader
that the  percolation theory corresponds to a theory of free
scalar field in Euclidean space-time.

The percolation theory introduces the  probability $p$
of a link to be ``occupied''. A connected sequence of occupied links is
called trajectory. The percolation theory is mapped into field theory
by the relation
\begin{equation}
p~\equiv~-\ln(M\cdot a)~,
\end{equation}
where $M$ is the bare mass and $a$ is the lattice spacing, $a\to 0$ in the
continuum limit.

The mass  gets renormalized and the physical mass
of the bosonic field is related to the bare mass
through the balance of action and entropy:
\begin{equation}\label{physical}
m^2_{\phys}~=~{const\over a}\Big[M(a)-{\ln 7\over a}\Big]~~,
\end{equation}
where the $\ln 7$ is specific for cubic lattice in 4d and we reserved for
dependence of the bare mass $M(a)$ on the lattice spacing.
Note that the physical mass $m^2_{\phys}$ is independent on the lattice spacing
$a$ only if there is almost exact cancelation between
the two terms in the right hand side of Eq. (\ref{physical}).
This is the action-entropy balance.

At a critical value, $p_c$ there appears an
infinite cluster of trajectories.
 In our case $p_c=1/7$. In the standard field-theory language this point corresponds to the physical mass (\ref{physical})
mass becoming tachyonic. The probability
of a given link to belong to the infinite cluster~is
\begin{equation}\label{percolation}
%\theta_{inf}
\theta_{\perc}
~\sim~ (p-p_c)^{\alpha}\,, \qquad \alpha~>~0\,.
\end{equation}
Moreover the emergence of the infinite cluster signals
the spontaneous symmetry breaking and
\begin{equation}
|\langle \phi \rangle|^2\sim
%\theta_{inf}
\theta_{\perc}\,,
\end{equation}
where $\phi$ is a Higgs field \cite{chernodub1}.
It is amusing that percolation produces the Higgs phenomenon
without an explicit introduction of the Higgs potential.
The interpretation of the lattice data at zero temperature
on the magnetic monopoles
in terms of the percolation theory can be found in \cite{chernodub1}.

\subsection{Monopole percolation around $T_c$: evidence for strings }

Percolating monopole trajectories fall on surfaces which are magnetic
strings. Quantum geometry of surfaces (or strings) is less developed.
Nevertheless, such concepts as action-entropy balance are well known
as well. Moreover, the monopole trajectories are strongly correlated with
the vortices and cover them densely. This observation allows us
to use alternatively the languages of the 1d and 2d defects.
In particular this trick (replacing the 2d defects by the 1d defects)
was used in \cite{chernodub} to argue, on the basis of the percolation theory,
that at $T>T_c$ the magnetic degrees of freedom become part of
the thermal Yang-Mills plasma. Let us review briefly the argumentation and
emphasize that in fact the lattice data cannot be understood in terms
of (effective) 4d field theory of monopoles. Instead, one should invoke
strings which are non-local objects.

One begins again [as in case of (\ref{physical})] with free field theory
in Euclidean time but now at non-zero temperature, $T\neq 0$.
One can then show that the so called wrapped trajectories which start
at Euclidean time $\tau=0$ and reach the other boundary $\tau=1/T$
(and then ``wrapped'' back because of the periodic boundary condition)
correspond to real (as opposed to virtual)  particles.
Intuitively, this is appealing since the wrapped trajectories visualize
particles which 'exist for ever'.
In terms of equations~\cite{chernodub}:
\begin{equation}\label{free}
\rho(T)~=~n_{\w}~~,
\end{equation}
where $n_{\w}$ is the density of the wrapped trajectories and
$$\rho(T)~\equiv~\int{{d^3{\bf p}\over(2\pi)^3}
{1\over [\exp(\omega_{\bf p}+\mu)/T]-1}}~~,$$
A crucial test of applicability of (\ref{free}) to the lattice monopoles
is provided by measuring how the density $n_{\w}$
scales with the lattice spacing. It should be in
physical units, independent of the lattice spacing. This, highly
non-trivial constraint is indeed satisfied by the data \cite{massimo}.

So far, there is similarity between standard thermal particles and
lattice monopoles. There exists, however, a crucial difference as well.
The point is that for ordinary scalar particles the trajectories
percolate uniformly in all four dimensions also at $T\neq 0$.
This is a generic feature of a local field theory
which does not depend, in particular, on the sign of $m^2_{\phys}$,
see Eq. (\ref{physical}).
The wrapped trajectories arise when a percolating trajectory is so to say
'trapped' by the periodic boundary condition and becomes a wrapped trajectory.
Thus, the Boltzmann factor corresponds to the density of virtual particles
at $T=0$ which propagate distance $1/T$.

The behavior of the trajectories of the lattice monopoles around
$T=T_c$ is very different. Namely,
both in case of 1d and 2d defects one can introduce a {\it local}
order parameter which quantifies the time alignment of the defects.
In case of monopoles  this quantity is~\cite{suzuki:A}:
\begin{equation}{\label{asymmetry}}
A~=~{3 \langle |n_t|^2 \rangle - \langle |{\bf n}|^2 \rangle \over ~{
3 \langle |n_t|^2 \rangle + \langle |{\bf n}|^2 \rangle }}\,,
\end{equation}
where the temporal, $|n_t|$, and spatial, $|{\bf n}|$, currents take values $1,0$
depending on the direction of the monopole trajectory (on the 4d cubic lattice).
In Eq.~\eq{asymmetry} a statistical average over all monopole configurations
is assumed.

In particular it was demonstrated in \cite{ejiri} that the asymmetry
(\ref{asymmetry}) measured on the trajectories
can serve as an
order parameter which equals zero in
the confined phase and is positive in the deconfined phase.
Let us emphasize again that the local asymmetry (\ref{asymmetry})
at points inside the lattice volume can arise because the monopoles
are in fact defects living on the surfaces, or strings. And the strings,
in turn, are non-local and can ``learn'' about the time direction because the
boundary condition in the time direction is singled out by its periodicity.

Similar asymmetry was observed directly
for the magnetic vortices as well \cite{langfeld}.
Moreover, the transition to almost complete time alignment is quite
fast so that (numerically) the percolation picture for the
vortices is completely changed in the temperature interval
$$0.8 \, T_c~<T<~1.1 \, T_c~.$$
At lower temperatures the percolation picture for the vortices
is four dimensional while at the upper edge the percolation occurs in three
spatial dimensions.

Thus, at $T>T_c$ the time dependence becomes (approximately) trivial
and the vortices can be reduced to their 3d projections which are
1d trajectories percolating in a 3d time slice. Since the original vortices
are closed the trajectories are closed as well.

\subsection{Higgs-field condensate}

Now we come to the central point of this section: at $T >T_c$
the continuum-theory description of the 2d dimensional surfaces
simplifies greatly and can be given in terms of {\it 3d field theory}.
The reason is that the strings become time-oriented \cite{langfeld}.
Their geometry in time direction becomes trivial and the strings
can be characterized entirely in terms of their 3d projection.
The 3d projections, or intersections of the 2d surfaces with a 3d
volume of a time slice represent 1d defects, or trajectories.
Finally, in quantum geometry, trajectories in any number of dimensions
can be mapped to a field theory.
%Thus, for the time-oriented surfaces
% we come to discuss
%3d field theories.
%As a result, their 3d projections to a given time slice become trajectories,
%or 1d defects in three-dimensional space. And 1d defects correspond to a field
%theory in any number of dimensions.

Thus, we are invited to interpret the lattice data on the magnetic
defects in terms of a 3d field theory.
In particular, it is crucial that the 1d trajectories considered do form an infinite cluster in a 3d time slice \cite{langfeld}.
The total length of the percolating cluster
can be defined in terms of the
density $\rho_s$ of
the percolating vortices in the three-dimensional timeslice:
\begin{equation}\label{good}
L_{\perc}~\equiv~\rho_s V_{3d}
\end{equation}
where $V_{3d}$ is the total 3d volume of the time slice.
The density $\rho_s$ was measured
and turned
proportional to the tension of the spatial Wilson loop $\sigma_s$:
\begin{equation}\label{densitym}
\rho_s~\sim~\sigma_s~ \sim ~ T^2 \, g^4(T) ,
\end{equation}
where $g^2(T)$ is the running coupling of the original 4d gauge theory,
and the combination $g^2_3(T) = T \, g^2(T)$ corresponds to the 3d gauge
coupling of the dimensionally reduced theory. The last relation in
Eq.~\eq{densitym} is valid  in the deconfined phase at temperatures
larger than $2 T_c$~\cite{bali}. The result (\ref{densitym}) is
non-trivial since there is no dependence on the lattice
spacing, as it should be for physical fields and would not be true for the
lattice ``artifacts''.

Combining the observation (\ref{densitym}) with Eq.~(\ref{percolation}) we come to the conclusion that
there exists a 3d scalar field $\Sigma_M$ condensed in the 3d vacuum:
\begin{equation}\label{analog}
\langle \Sigma_M\rangle ~\neq ~0\,.
\end{equation}
A reservation is that (\ref{analog}) assumes that the time orientation
of the magnetic strings is complete and the 3d description is adequate.
Near to $T=T_c$ this picture should be taken with caution since
there are still ``wiggles'' of the surfaces in the time direction.
In the 3d projection such wiggles would be manifested as
breaking/disconnection
of the infinite cluster. Instead of an infinite cluster one would
observe rather a few 'big' clusters. With temperature going up
such disconnections
should occur more rarely.

\subsection{Higgs-field quantum numbers }

Thus, lattice provides a definite evidence in favor of a 3d magnetic Higgs
field at $T>T_c$. Similar prediction arises in the dual models \cite{gorsky}.
It is only on the lattice, however, that the prediction can be directly checked.
There are limitations to the use of the lattice of course as well.
In particular, one would like to learn more on interactions of the field
$\Sigma_M$. The lattice data are limited, however to the magnetic degrees of freedom.
Lattice is 'blind' (so far) to ``electric degrees of freedom'', say, to the
same Polyakov's lines (see the Introduction). According to the string picture,
non-vanishing vacuum expectation value of the Polyakov line, $\langle L \rangle \neq 0$
corresponds to 3d scalar particles in the adjoint representation, $\Pi^a$.
Such particles are not visible directly on the lattice
with the presently available techniques.

It is amusing that -- despite of all these apparent limitations -- we can still
state that the $\Sigma_M$ field has a negative parity (in case of
the original SU(2) Yang-Mills theory). Indeed, introduce phenomenologically possible
interaction terms:
\begin{eqnarray}\label{phen}
L_{int}~\sim~s_3\Sigma_M^3~+~s_4\Sigma_M^4~+~s_5\Sigma_M^5~+~\\ \nonumber s_{1,2}\Sigma_M(\Pi^a)^2~
+~s_{2,2}\Sigma_M^2(\Pi^a)^2+...
\end{eqnarray}
Then it is known that terms with an odd number of $\Sigma_M$-legs is not allowed:
\begin{eqnarray}
s_3~=~s_5~=s_{1,2}~=~0~~.
\end{eqnarray}
Indeed, the vertices corresponding to the interactions (\ref{phen})
would be seen on the lattice as
intersections between the odd number of the corresponding
trajectories\footnote{Such intersections are observed for the monopole
trajectories at T=0, see \cite{chernodub1} and references therein.}.
However, the trajectories of the magnetic strings are closed and the odd-number
intersections are impossible.

In terms of the interactions (\ref{phen}) one can say that the $\Sigma_M$-field
has a negative parity and the interaction Lagrangian is invariant under this
parity transformation. We will return to this point later.

\subsection{Matching 3d theories of confinement}

The 3d sector of Yang-Mills theories at high temperature is
a confining theory in the sense that the spatial Wilson line obeys
area law at all temperatures. Thus, we expect that if the emerging
picture of the magnetic component in 3d is correct it should also
produce understanding of this area law. The problem is currently
under investigation \cite{henri}.
The preliminary finding is that, indeed, the properties of the
$\Sigma_M$-field do match models of the type \cite{thooft1}.

\subsection{What to measure next}

One could get further information on the properties
of the $\Sigma_M$ field from lattice measurements.

An interesting quantity to measure is the non-Abelian action associated
with the 1d defects. The action-entropy balance assumes that
\begin{equation}
(\mathrm{Action)}_{1d}~\approx~ {\ln 5\over a}L~~,
\end{equation}
where $L$ is the length of trajectory. In case of four Euclidean dimensions the factor $\ln 5$ is replaced
by $\ln 7$ and the latter relation was confirmed for the lattice monopoles
at T=0 \cite{bornyakov1}.

The mass of the excitations of the field $\Sigma_M$ can be inferred
from measuring distribution of finite cluster in length:
\begin{equation}
N(L)~\sim~\exp (-m^2_{\Sigma}\, L\, a)\,.
\end{equation}
Furthermore, measuring extra action associated with self-intersections
of the trajectories would allow to measure the coefficient $s_4$,
see Eq. (\ref{phen}).

%Consider now lattice realistic lattice simulations at finite temperature $T$.
%The lattice spacing $a$ is related to temperature via the relation
%$a = 1/(N_t T)$, where $N_t$ is the (fixed) number of the lattice steps in the
%temperature Euclidean direction. The Higgs condensate is proportional
%to the density of the percolating cluster~~\cite{chernodub1},
%$\langle \Sigma_M\rangle^2 \sim a \rho_s$. Using the relation
%$a \sim 1/T$ and Eq.~\eq{densitym} we then get the following prediction for the
%expectation value of the condensed Higgs field on the lattice with at fixed $N_t$:
%\begin{equation}\label{newcondensate}
%\langle \Sigma_M\rangle^2~\sim~T^2\cdot a\cdot g^4(T)~\sim~T\cdot g^4(T)\,.
%\end{equation}

\section{Three dimensional models with $Z_2$ parity}

Now we will review briefly an independent approach to the theory of the deconfinement which
started with effective theories for the Polyakov's line [see (\ref{potential})]
and developed into superrenormalizable 3d  field theories,
see \cite{yaffe,pisarski,kurkela} and references therein.
In particular we will compare the 3d theory of Ref. \cite{kurkela}
with the models favored by the lattice data on the magnetic component,
as discussed in the preceding section.

The model \cite{kurkela} postulates existence of a 3d color field $\Pi^a$ and
of a colorless field $\Sigma_P$ with following potential energy:
\beqn
\label{new}
V (\Sigma_P,\Pi_a ) & = & b_1 \Sigma_P^2 + b_2 \Pi_a^2
\\
& & + c_1\Sigma_P^4 + c_2 (\Pi_a^2)^2 + c_3 \Sigma_P^2 \Pi_a^2\,.
\nonumber
\eeqn
While the field $\Pi^a$ can be identified with the original filed $A_0^a({\bf x})$,
the $\Sigma_P$-field is a new degree of freedom. The introduction of this new field turns
to be crucial to correctly reproduce the thermodynamics of the original 4d theory
in terms of the effective 3d theory (\ref{new}). It is amusing that
numerically it is indeed possible to match the 3d theory to its original 4d counterpart
beginning practically with $T\ge T_c$. At high temperature, one assumes that the new field $\Sigma_P$ can be integrated out to reproduce the standard dimensionally
reduced theory. To justify this procedure one assumes that the excitations of the
$\Sigma_P$ field are heavier (by an inverse power of the coupling $g(T)$)
than excitations of the field $\Pi^a$.

Although potential (\ref{new}) might look somewhat arbitrary it is motivated by
remarkably simple symmetry considerations. One can say that the form (\ref{new}) is
the minimal realization of these symmetries possible.
Indeed, the basic properties of (\ref{new}) is that it is invariant under
$$\Sigma_P~\to~-\Sigma_P$$
and that $\langle \Sigma_P\rangle\neq 0$. To ensure these properties one needs
nonvanishing coefficients $b_1,c_1$. The $(\Pi^a)^2$ and $(\Pi^a)^4$ terms
are standard for any dimensional reduction of the original Yang-Mills theory.
Finally, the $\Sigma_P^2(\Pi^a)^2$ interaction is needed
for the $\Sigma_P$ field not to be decoupled from the rest of the system.

The central point is that the invariance under the reflection
$\Sigma_P\to-\Sigma_P$ and the fact that $\langle \Sigma_P\rangle \neq 0$ both corresponds
the properties of the
underlying 4d Yang-Mills theory. Namely, the $\Sigma_P$ field represents
the vacuum expectation value of the
Polyakov line (and its fluctuations) at $T>T_c$. The parity transformation
$\Sigma_P \to -\Sigma_P$ is then an analogue of the $Z_2$ transformation of the
original Polyakov line.

\section{Comparison of the two approaches}
\subsection{Thermodynamics of plasma}

The magnetic component on the lattice appears to be
crucial for thermodynamics of the whole plasma
at temperatures about $T_c\le T\le 3T_c$.
This follows from measurements of the contribution
of the 4d vortices into the equation of state of the Yang-Mills
plasma \cite{nakamura}. In the language of the 3d models a qualitatively
similar conclusion was made in Ref. \cite{kurkela},
this time about the role of the newly introduced Higgs field $\Sigma_P$.

It is amusing that both the field $\Sigma_P$ and the magnetic
component in the 4d language are tachyonic. In case of the
vortices the contribution of the vortices to the energy and
pressure densities is of unphysical, negative sign.
The negative sign corresponds to the ``negative number'' of degrees
of freedom, which are balanced by the rest of the plasma.
The $\Sigma_P$-field, on the other hand,
condenses in the vacuum and is tachyonic in this sense as well.
Unfortunately, more detailed comparison is not possible, since in case
of the vortices the sign is defined with respect to $T=0$. In case
of the $3d$ models  such a normalization is not possible
since these models do not work and not defined at $T=0$.

Note that in both cases (4d vortices and 3d Higgs fields) the  colorless
scalar field
is not the only important ingredient of the theory. In case of
the vortices this follows from measuring contribution to the equation of
state from the ``rest of plasma''. In case of the $\Sigma_P$-field, it is
explicitly one of four Higgs fields strongly interacting with each other.

\subsection{Limitations of the 3d picture}

In both approaches, we do not expect that the 3d picture remains valid
in the continuum limit of the lattice spacing going to zero,
$a\to 0$ with $a\ll 1/T$.

In case of theories which start with the Polyakov line the subtle point
is the ultraviolet divergence
associated with the operator $L$. This is the self-energy linear
divergence common to all the Wilson lines on the quantum level:
$$\langle L\rangle~\sim~\exp\big(-M(a)\cdot T\big)\,,
\qquad
M(a)~\sim~g^2/a~~,$$
where $a$ is the lattice spacing.
To avoid this divergence one does not allow to consider the limit
$a\to 0$ or, in another language, one considers smearing
(``coarsing'')
of  Polyakov line  over transverse finite distances of order 1/T.
In particular,  Ref. \cite{kurkela} assumes the following coarsing:
\begin{equation}\label{coarse}
L(x)~ \rightarrow~ {1\over V_{\mathrm{block}}}\int d^3 x\,   U (x, y) L(y)U (y, x),
\end{equation}
where the integration goes over the (somewhat arbitrary) $O(T^{3})$
volume of the block and $U (x, y)$ is a Wilson line connecting
the points $x$ and $y$ at constant time $\tau = 0$.

Necessity for such a smearing makes direct contact with the continuum
limit $a\to 0$ difficult and introduces  uncertainties in understanding
dynamics of the Polyakov lines.
To our mind, the assumption that the UV divergence is smeared out is
equivalent to the hypothesis that there exist electric
strings with {\it physical} tension $\sigma~\sim~\Lambda_{QCD}^2$
at $T=0$. As it is
common in quantum geometry, the physical tension is
a result of cancelation between UV divergent (bare) action and entropy:
$$ \sigma\cdot(\mathrm{area})~\sim~(\mathrm{action})~-~(\mathrm{entropy})\,.$$
Both the action and entropy are of order $(area)/a^2$, where the area
corresponds to the worldsheet of the string in the Euclidean space-time.

In case of the geometrical picture, or magnetic defects some coarsing is
also implied. To see this, consider  again the vacuum expectation value $\langle \Sigma_M\rangle$.  We discussed the lattice data which show that the density $\rho_s$ is in physical
units, see (\ref{good}). However, in the continuum limit one has~\cite{chernodub1}:
\begin{equation}\label{nogood}
{\langle \Sigma_M\rangle}^2~\sim~\rho_s\cdot a~~,
\end{equation}
where $a$ is the lattice spacing.
By sending $a\to 0$ naively we would get $\langle \Sigma_M\rangle = 0$.
But this would be true if the $\Sigma_M$ were a fundamental field of
a genuine 3d theory.
In case of an effective theory
we may assume $a\sim T^{-1}$ so that the condensate is expressed
in physical (not in the lattice) units.

At very short distances, or very small lattice spacings $a$
we expect to see again the 4d percolation
picture of the vortices. Indeed, observation of a percolating
cluster implies {\it action-entropy balance} at short
distances, for details see \cite{polyakov3,chernodub1}.
The precise form of the balance depends on the number of dimensions.
The vortices are 4d objects and at short distances one expects
restoration of the 4d picture, with no infinite, percolating cluster
in the 3d projection\footnote{In case of the vortices such a phenomenon
has not been observed yet on the lattice. As we believe, this happens because
one cannot perform lattice simulations with small enough lattice spacings.
However, a similar phenomenon in case of the magnetic monopoles has been
observed recently, M. D'Elia and A. D'Alessandro, private communication.}.

To summarize, the use of the 3d language for the Higgs fields is justified
in the infrared regime:
provided a finite lattice spacing, or the coarsening scale is introduced
for the {\it spatial} dimensions\footnote{This relation can be fulfilled for the lattices with space-time asymmetric spacing.
Such asymmetry may be achieved by utilizing different lattice couplings spatial and temporal directions.},
$a\ge 1/T$.
This is true both from the continuum- and lattice-theories
perspective. A delicate point is that a priori it is not clear that
this reservation can be indeed implemented because the time extension
is also of order $\tau_{\max}\sim 1/T$. Success reported in Ref. \cite{kurkela}
supports the hypothesis that numerically the scheme still works.
In the lattice version of the magnetic component of the plasma the same
conclusion could be in fact made on the basis of the data \cite{langfeld}.
Indeed, the behavior $\rho_s\sim T^2$ was observed for a range of the lattice
spacings, although at very small spatial lattice spacing $a$ it is expected to break down.
However, we expect that the description of the infrared long distance 3d physics in terms of the
condensed (magnetic) scalar field(s) should be a useful tool for study of the nonperturbative
properties of the high temperature phase.

\subsection{Symmetries involved}

The symmetry behind introducing the $\Z_2$ parity of the $\Sigma_P$ field
is the global $\Z_2$ invariance of the original Yang-Mills theory
in the lattice formulation. Let us remind the reader definition of this symmetry.
In the discretized, lattice formulation the plaquette action $S_{x,\mu\nu}$
is given in terms of unitary matrices $U_{x\mu}$:
\begin{equation}\label{latticeaction}
S_{x,\mu\nu}~\sim~{\mathrm{ Tr}} \big(U_{x\mu} U_{x+\hat\mu,\nu} U_{x+\hat\nu,\mu}^{\dagger} U_{x\nu}^{\dagger}\big)~,
\end{equation}
where
\begin{equation}
U_{x\mu}~\sim~\exp\Big\{ia \hat{A}_{\mu}(x) \Big\}\,,
\end{equation}
$\hat{A}_{\mu}(x)$ is the discretized vector potential
and $a$ is the lattice spacing. The continuum action
arises by expanding the lattice matrices $U$ in series of the lattice spacing,
\beqn
U \approx \bbbone + i a \hat{A}_{\mu}(x) + O(a^2)\,.
\eeqn
The global symmetry in point is the change of the sign of
all the matrices $U$ which correspond to links in the time direction at
a given time, $\tau=0$. The lattice action is not changed since the signs of two
matrices $U$ are changed simultaneously.

Note that after the change of the sign the matrices $U$ are close
to the matrix $-\bbbone$ while the standard continuum limit corresponds to the
matrices $U$ in the vicinity of the unit matrix, $+\bbbone$.
It is worth emphasizing  that existence of the $\Z_2$ symmetry
is not automatic for any discretization. In particular, there is no
$\Z_2$ symmetry if the discretized potential $\hat{A}_{\mu}\equiv A_{\mu}^a\, t_a$
is written in terms of the matrices $t_a$ in the adjoint representation
of the gauge group rather than in the fundamental representation as we assumed above.
The continuum-limit action remains however the same.

Since the $\Z_2$ global symmetry is a property of  a
particular regularization scheme, one would expect
that thermodynamical properties do not depend on existence
of the $\Z_2$ symmetry. On the other hand, the whole of the message of
the Ref.~\cite{kurkela} is that introduction of the $\Z_2$ parity is crucial
to reproduce the thermodynamics of the original theory, that is
the second order phase transition.

Now, in the geometrical language of the defects we do not
introduce any $\Z_2$ parity but account for the fact
that there exist magnetic degrees of freedom, or defects.
Existence of such defects seems to be a generic feature of the Yang-Mills theories
formulated in the continuum language.
In particular, holographic formulations do imply
existence of the magnetic defects.
Reduction of the magnetic defects to a $3d$ scalar, $\Sigma_M$  field is,
again, a generic feature of these models~\cite{gorsky}.

\section{Conclusions}

We have argued that the lattice data provides strong
evidence that in the deconfining phase
the magnetic
vortices reduce to their 3d projection
and can be described as a color singlet 3d Higgs field, $\Sigma_M$
which is condensed in the 3d vacuum. This picture holds
in the ``infrared regime'' (provided that the {\it spatial} resolution of the measurements
is not too fine, that is $a\ge 1/T$ at high temperatures).
The observations agree with the dual models mentioned in the Introduction.
Indeed strings (or D-branes) become tensionless
in the deconfining phase only if they are wrapped around the periodic
Euclidean time direction. This is the explanation
in the language of the dual models why our surfaces are aligned with
the time direction. Moreover, reduction of the strings to 3d particles
follows from T-duality.

There are strong similarities between the properties of
the magnetic scalar field $\Sigma_M$ which is introduced in this paper,
and the 3d Higgs field $\Sigma_P$ which was recently introduced for absolutely
independent reasons, see \cite{kurkela}. This similarity
supports validity of the both pictures in their gross features.

Despite of these similarities of their properties, the symmetries behind
the emerging fields $\Sigma_M$ and $\Sigma_P$ are different.
The reason for the introduction the field $\Sigma_P$ is $\Z_2$ symmetry which
is a property of a particular ultraviolet regularization of Yang-Mills
theories. Appearance of the field $\Sigma_M$ reflects the fact that
generically there exist both electric and magnetic defects.

Generally speaking, one cannot rule out  that the family of the 3d scalars
includes both $\Sigma_M$ and $\Sigma_P$. Indeed, these fields were
introduced for very different reasons. The most economic version which
identifies $\Sigma_M$ and $\Sigma_P$ seems more attractive since in this case
the number of the scalars agrees with predictions of the dual model.

Finally, the magnetic component of the Yang-Mills \cite{chernodub,shuryak,sasha,nakamura,massimo}
plasma gets a novel interpretation in terms of the 3d
field theoretic models. This interpretation may provide us with a great simplification
of the treatment of the gluonic plasma compared to the strings in original four-dimensional theory
which hosts the original image for the magnetic component of the plasma.

\acknowledgments

This work was supported by Grants-in-Aid for Scientific
Research from ``The Ministry of Education, Culture,
Sports, Science and Technology of Japan'' Nos. 17340080
and 20340055, by the STINT Institutional grant IG2004-2 025,
by the grants RFBR 08-02-00661-a, DFG-RFBR 436 RUS, by a grant for scientific
schools NSh-679.2008.2, by the Federal Program of
the Russian Ministry of Industry, Science and Technology
No. 40.052.1.1.1112 and by the Russian Federal Agency
for Nuclear Power. 
V.I.Z appreciate his fruitful stays at Research Institute
for Information Science and Education of Hiroshima University,
Japan and at the Laboratoire de Mathematiques et
Physique Theorique at Universit\'e Francois Rabelais, Tours, France.

%\section*{References}

\end{document}